\documentclass[superscriptaddress,preprintnumbers,amssymb,nofootinbib,twocolumn]{revtex4-2}

\usepackage{bm}
\usepackage{esvect}
\usepackage{verbatim}
\usepackage{multirow}

\usepackage[left]{lineno}
\usepackage{blindtext}
\pdfoutput=1
\usepackage[T1]{fontenc}
\usepackage{graphicx}
\usepackage{epsfig}
\usepackage{amsmath}
\usepackage{amsfonts}
\usepackage{amssymb}
\usepackage{braket}
\usepackage{cancel}
\usepackage{pstricks}
\usepackage{color}
\usepackage{xcolor}
\usepackage{feynmf}
\usepackage[normalem]{ulem}
\usepackage{epstopdf}
\usepackage{soul}
\usepackage{cancel}

\usepackage[footnote]{witharrows}
\usepackage{braket}
\usepackage{cancel}
\usepackage{setspace}
\usepackage{pstricks}
\usepackage{xcolor}
\usepackage{float}
\usepackage{mathtools}
\usepackage{multirow}
\usepackage{booktabs}
\usepackage{enumitem}

\newcommand{\abs}[1]{\vert #1\vert}

\newcommand{\MeV}{{\rm MeV}}
\newcommand{\GeV}{{\rm GeV}}
\newcommand{\TeV}{{\rm TeV}}
\newcommand{\m}{{\rm m}}
\newcommand{\cm}{{\rm cm}}

\newcommand{\cL}{\mathcal{L}}

\newcommand{\cO}{\mathcal{O}}

\newcommand{\NPoT}{N_{\text{PoT}}}

\newcommand{\eg}{\textit{e.g.}}

\begin{document}

\preprint{IRMP-CP3-23-11, MPP-2023-40, KEK-TH-2499}

\title{Probing long-lived axions at the KOTO experiment}

%%%%%%%%%%%%%%%%%%%%%%%%%%%%%%%%%%%%%%%%%%%%%%%%
\author{Yoav Afik}
\email{yoavafik@gmail.com}
\affiliation{Enrico Fermi Institute, University of Chicago, Chicago, Illinois 60637, USA}
\affiliation{Experimental Physics Department, CERN, 1211 Geneva, Switzerland}

\author{Babette D\"obrich}
\email{babette.dobrich@cern.ch}
\affiliation{Max-Planck-Institut  für Physik (Werner-Heisenberg-Institut), F\"ohringer Ring 6, 80805 M\"unchen, Germany}

\author{Jan Jerhot}
\email{jan.jerhot@cern.ch}
\affiliation{Centre for Cosmology, Particle Physics and Phenomenology (CP3), Universit{\'e} Catholique de Louvain, Chemin du Cyclotron 2,
B-1348 Louvain-la-Neuve, Belgium}

\author{Yotam Soreq}
\email{soreqy@physics.technion.ac.il}
\affiliation{Physics Department, Technion--Institute of Technology, Haifa 3200003, Israel}

\author{Kohsaku Tobioka}
\email{ktobioka@fsu.edu}
\affiliation{Department of Physics, Florida State University, Tallahassee, Florida 32306, USA}
\affiliation{High Energy Accelerator Research Organization (KEK), Tsukuba 305-0801, Japan}
%%%%%%%%%%%%%%%%%%%%%%%%%%%%%%%%%%%%%%%%%%%%%%%%

\date{\today}

%%%%%%%%%%%%%%%%%%%%%%%%%%%%%%%%%%%%%%%%%%%%%%%%
\begin{abstract}
While the main goal of the J-PARC KOTO experiment is to measure the rare decay $K_L \to \pi^0 \nu \bar \nu$, the unique setup of KOTO raises the possibility to search for physics beyond the Standard Model, in an attempt to probe parts of the parameter space which are not covered by other experiments. In this paper, we test the possibility of using KOTO to search for heavy QCD axions, or axionlike particles, a well-motivated extension of the Standard Model emerging in a variety of models. In particular, we estimate the sensitivity of the current KOTO setup as well as KOTO Step 2 for various benchmark scenarios of axion coupling to the Standard Model. We find that KOTO Step 2 can probe new regions in the parameter space, while KOTO with its current form can only reaffirm the existing bounds.
The obtained axion datasets are available as an update of the public code of the {\sc{Alpinist}} framework, including implementation of KOTO setups in the simulation, allowing for interpretation of various analyses as searches for axions in custom models.
\end{abstract}
%%%%%%%%%%%%%%%%%%%%%%%%%%%%%%%%%%%%%%%%%%%%%%%%

%%%%%%%%%%%%%%%%%%%%%%%%%%%%%%%%%%%%%%%%%%%%%%%%
\maketitle
\flushbottom
%%%%%%%%%%%%%%%%%%%%%%%%%%%%%%%%%%%%%%%%%%%%%%%%

%%%%%%%%%%%%%%%%%%%%%%%%%%%%%%%%%%%%%%%%%%%%%%%%
\section{Introduction}
\label{eq:intro}
%%%%%%%%%%%%%%%%%%%%%%%%%%%%%%%%%%%%%%%%%%%%%%%%

Rare kaon decays are a well-known test of the Standard Model~(SM) and serve as a very sensitive probe of new physics~(NP). Two golden channels are $K^+\to \pi^+\nu\bar{\nu}$ and $K_L\to \pi^0 \nu\bar{\nu}$, which are very rare decays with a branching ratio (BR) at the $\sim10^{-11}$ level. The NA62~\cite{NA62:2017rwk} and KOTO~\cite{Yamanaka:2012yma} experiments aim at measuring these BRs for the first time, with the latest results given in Refs.~\cite{NA62:2021zjw,KOTO:2020prk}. 
Besides testing the SM, including the Grossman-Nir bound~\cite{Grossman:1997sk}, these measurements probe feebly interacting particles~(FIPs) which contribute to the $K\to \pi + \,{\rm invisible}$ decay see, \eg{} the recent review in Ref.~\cite{Goudzovski:2022vbt}.  

Both KOTO and NA62 are based on proton-fixed targets, with $30\,\GeV$ and $400\,\GeV$ beams, respectively, and far detection systems. They can effectively serve as beam dump experiments probing NP without relying on kaon decays, since NP particles can be produced already in the target. This was pointed out in Ref.~\cite{Dobrich:2015jyk} in the context of NA62 by using a special running mode of the experiment; see also Ref.~\cite{Dobrich:2019dxc}. This beam dump potential was pointed out as a possible explanation for the three candidate events in the KOTO 2019 data~\cite{Kitahara:2019lws}.

\begin{figure*}[t]
\begin{center}
\includegraphics[width=0.65\textwidth]{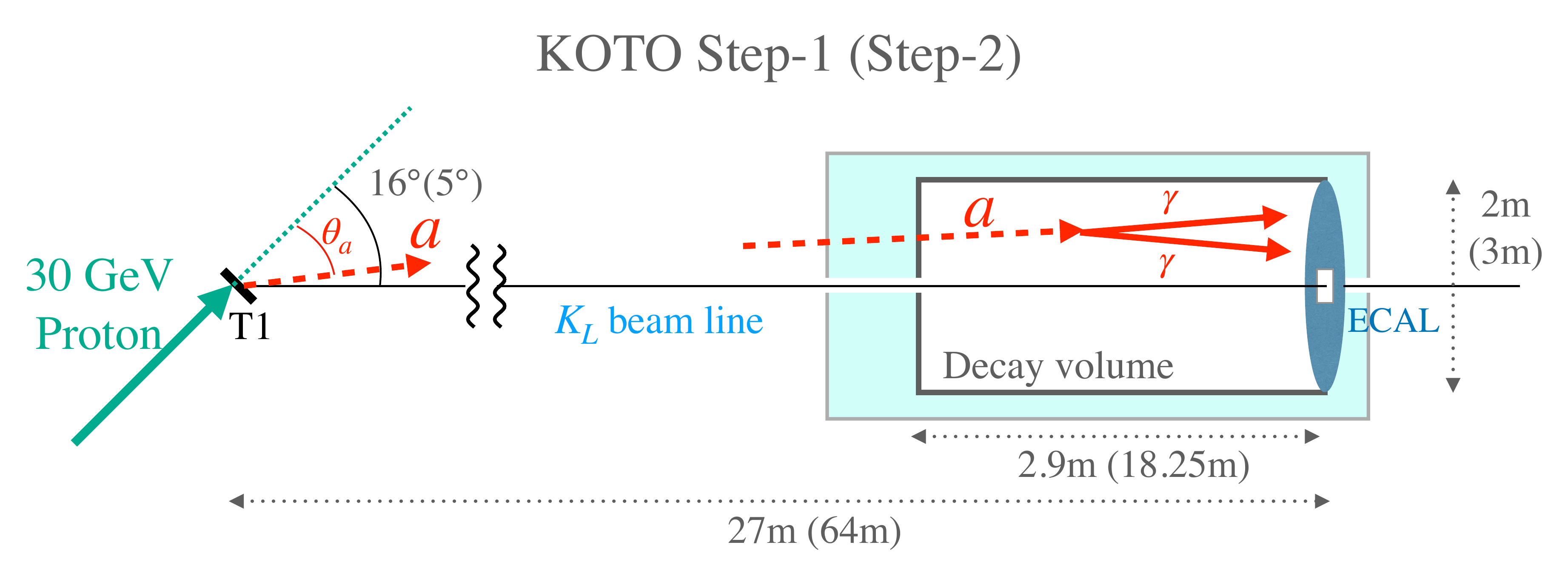}
\end{center}
\vspace{-7mm}
\caption{A schematic illustration of the KOTO layout for Step 1\,(Step 2). See text for details. }
\vspace{-2mm}
\label{fig:illustration}
\end{figure*}

In this work, we study the potential of the KOTO experiment to serve as a proton beam dump for sub-GeV NP searches. Unlike NA62, KOTO can probe long-lived new particles in a digamma final state during its kaon physics running without needing a dedicated trigger or run, in a completely parasitic scheme; see Fig.~\ref{fig:illustration}. This limitation is caused by the NA62 trigger requiring the presence of a kaon in the standard data-taking. Due to the differences in the beam energy and other geometrical factors, we expect that KOTO will explore a different region of the parameter space than NA62 and past proton beam dump experiments. In particular, we show that the future run of KOTO can search for NP particles, which are associated with solutions to the strong $CP$ problem.

%%%%%%%%%%%%%%%%%%%%%%%%%%%%%%%%%%%%%%%%%%%%%%%%%%%%%
%%%%%%%%%%%%%%%%%%%%%%%%%%%%%%%%%%%%%%%%%%%%%%%%%%%%%

Our primary benchmark model is the axion, $a$, which is a compelling addition to the SM because it potentially solves the strong $CP$ problem~\cite{Peccei:1977hh,Peccei:1977ur,Weinberg:1977ma,Wilczek:1977pj}, and it can be by itself a dark matter~(DM) candidate~\cite{Preskill:1982cy, Dine:1982ah, Abbott:1982af}. A similarly motivated case is the axionlike particle~(ALP) as a pseudo-Nambu-Goldstone boson of spontaneously broken global symmetry at a high scale $f_a$. It can also provide a portal to the dark sector~\cite{Nomura:2008ru,Freytsis:2010ne,Dolan:2014ska,Hochberg:2018rjs}. In both cases, we focus on the SM gauge field interactions, which are given by
\begin{align}
    \cL \supset
    &\, c_{GG} \frac{ \alpha_s a}{8\pi f_a}  G^{a}_{\mu\nu}\tilde G^{a \mu\nu}
    +c_{BB} \frac{ \alpha_Y a}{8\pi f_a}  B_{\mu\nu}\tilde B^{\mu\nu}\nonumber\\
    &+c_{WW} \frac{ \alpha_2 a}{8\pi f_a}  W_{\mu\nu}\tilde W^{\mu\nu} \, ,
\end{align}
where $c_{GG,BB,WW}$ are dimensionless parameters. The SM gauge field strength is given by $G^a_{\mu\nu}$, $B_{\mu\nu}$ and $W_{\mu\nu}$ for the strong, hypercharge, and weak interactions, respectively; $\alpha_s=g_s^2/(4\pi)$ is the strong gauge coupling, and similarly for $\alpha_Y$ and $\alpha_2$. We assume the axion/ALP mass is sub-GeV, heavier than the QCD contribution in the light of heavy QCD axion models~\cite{Fukuda:2015ana, Agrawal:2017eqm, Agrawal:2017ksf, Gaillard:2018xgk, Gherghetta:2020keg,Gupta:2020vxb, Gherghetta:2020ofz, Valenti:2022tsc}, which revives the low decay constant from the long-standing bounds~\cite{CAST:2017uph, Raffelt:2006cw, Raffelt:1996wa, Friedland:2012hj, CHARM:1985anb, E949:2005qiy}. Furthermore, $f_a\lesssim 10\,\TeV$ is  favored by the axion quality problem~\cite{Agrawal:2017ksf}. Since this scenario is potentially discovered in laboratories, experimental data have been reinterpreted, leading to additional constraints. Promising probes based on future experiments have also been proposed~\cite{Bjorken:1988as, Blumlein:1990ay, CHARM:1985anb, E949:2005qiy, NA62:2021zjw, Bauer:2017ris, Aloni:2018vki, Hook:2019qoh, Kelly:2020dda,  Chakraborty:2021wda, Bertholet:2021hjl, Bauer:2021wjo}. Hereafter, we collectively refer to the heavy QCD axion and ALPs as axions, where the mass and the couplings are independent parameters. 
    
In the following, we discuss the KOTO experimental setup and data-taking modes in Sec.~\ref{sec:setup}. In Sec.~\ref{sec:flux}, we describe the axion production and decay. The quantitative impact of this analysis is shown in Sec.~\ref{sec:limits}, where we derive the bounds from current data and estimate the projection for future data-taking. We conclude in Sec.~\ref{sec:conclusion}.

%%%%%%%%%%%%%%%%%%%%%%%%%%%%%%%%%%%%%%%%%%%%%%%%
\section{KOTO setup and data-taking}
\label{sec:setup}
%%%%%%%%%%%%%%%%%%%%%%%%%%%%%%%%%%%%%%%%%%%%%%%%

In this work, we exploit several past and future setups of the KOTO experiment while accounting for the available information on the experimental conditions. These setups fall into two independent categories: one regarding the experiment layout and one regarding the data-taking mode.

We consider two experimental layouts as follows:
\begin{itemize}
\item Step 1: The present 2022 layout, which was also used in the 2015 data-taking~\cite{KOTO:2018dsc}.
\item Step 2: The proposed setup for the future as described in Ref.~\cite{Aoki:2021cqa}.
\end{itemize}
A schematic view of the setups is found in Fig.~\ref{fig:illustration}. For both steps we consider two data-taking modes: 
\begin{itemize}
\item Kaon mode: The standard mode with a $K_L$ beam.
\item Beam-dump mode: A special run in a beam dump mode, which includes a shield that blocks the beam (beam plug) and different selection cuts. 
\end{itemize}
% 

%%%%%%%%%%%%%%%%%%%%%%%%%%%%%%%%%%%%%%%%%%%%%%%%
\subsection{The experimental setups}
\label{sec:experimentalsetups}
%%%%%%%%%%%%%%%%%%%%%%%%%%%%%%%%%%%%%%%%%%%%%%%%

In both setups, the experiment uses a primary $30\,\GeV$ proton beam from the J-PARC main ring. The proton beam impinges on a golden target T1 and generates a secondary hadronic beam which, besides other particles, consists of $K_L$. In the present setup, Step 1~\cite{KOTO:2018dsc}, the experiment axis is under a $16^\circ$ angle with respect to the primary proton beam, and its front end is located $21\,\m$ from the T1 target with a set of collimators, sweeping magnets in between for forming the neutral $K_L$ beam, and veto detectors for upstream background suppression. The CsI calorimeter (ECAL), located $27\,\m$ from the T1 target, has a $2\,\m$ diameter with a $15\,\cm\times 15\,\cm$ central hole for the beam. The decay volume is $2.9\,\m$ long and precedes the ECAL.

The proposed KOTO Step 2 setup~\cite{Aoki:2021cqa} with a higher intensity beam assumes a $5^\circ$ angle between the detector and the primary beam. We assume the beginning of the decay volume to be $45.75\,\m$ away from the target. The calorimeter size is increased to a $3\,\m$ diameter with a $20\,\cm\times 20\,\cm$ central hole for the beam and it is located $64\,\m$ from the target.

In a special beam dump mode during the operation with the Step 1 setup~\cite{KOTO:beamdump}, a beam plug was placed to close the $K_L$ beamline, however, the sweeping magnet was not functional. The dataset of this run is smaller than in the kaon mode. We do not know if the backgrounds stated in Ref.~\cite{KOTO:beamdump} could be further reduced at the analysis level, but without a functioning sweeping magnet, a 0-background setting with the acquired data seems unlikely to us. With a functional sweeping system or a dedicated run with optimized magnet sweeping, one may be optimistic that a small background can be achieved.

The above-described experimental layouts and modes are implemented in the {\sc Alpinist} framework~\cite{Jerhot:2022chi} together with specific selection conditions, such that simplified simulations of axion production and decays can be performed for the interpretation of KOTO sensitivity for axion detection. The updated code and datasets are publicly available at \url{https://github.com/jjerhot/ALPINIST}. Details of the datasets and selection conditions are given in the following Sec.~\ref{sec:data}.

%%%%%%%%%%%%%%%%%%%%%%%%%%%%%%%%%%%%%%%%%%%%%%%%
\subsection{KOTO data-taking modes and their interpretation}
\label{sec:data}
%%%%%%%%%%%%%%%%%%%%%%%%%%%%%%%%%%%%%%%%%%%%%%%%

While operation in the beam dump mode can potentially allow a direct search for particles beyond the SM in a background-clean environment, the majority of the data are collected in the kaon mode, with the main aim to measure the extremely rare $K_L\to\pi^0\nu\bar{\nu}$ decay.

%%%%%%%%%%%%%%%%%%%%%%%%%%%%%%%%%%%%%%%%%%%%%%%%
\subsubsection{Kaon mode}
\label{sec:Kaon-mode}
%%%%%%%%%%%%%%%%%%%%%%%%%%%%%%%%%%%%%%%%%%%%%%%%

In the kaon mode, searching for particles of different origin than in the $K_L$ decay may prove to be challenging due to the backgrounds originating both from the beam and from upstream. Nevertheless, since $K_L \to \pi^0 \nu \bar{\nu}$ with a consequent $\pi^0 \to \gamma\gamma$ decay has the same signature as the axion $a \to \gamma\gamma$ decay, this provides an opportunity to reinterpret the (non)observation of the $\pi^0\nu\bar{\nu}$ signal to constrain also the axion parameter space, see \eg Ref.~\cite{Kitahara:2019lws}. Unlike SM particles, axions can propagate through the beamline elements and upstream detectors and they can decay off-axis from the neutral $K_L$ beam. Obviously, the different kinematics of these two options would render a dedicated analysis for the case of the axion possibly more sensitive than what we state below.

For the reinterpretation of the $K_L\to\pi^0\nu\bar{\nu}$ search as an axion search, we reanalyze the $a\to \gamma\gamma$ decays simulated using the toy Monte Carlo implemented in the {\sc Alpinist} framework, assuming the $K_L \to \pi^0 \nu\bar{\nu}$ selection conditions for Step 1 analysis of 2015 data~\cite{KOTO:2018dsc}, for the future Step 1 run~\cite{KOTO:ichep22} and for Step 2~\cite{Aoki:2021cqa}. These are summarized in Table~\ref{tab:KaonModesSelections}. In particular, we are implementing cuts on the following kinematic variables: the photon cluster coordinates in the plane perpendicular to the $K_L$ beam axis, $x_{\gamma_{1,2}}$ and $y_{\gamma_{1,2}}$; and $r$, which is the separation distance between the photon clusters. $R_{\text{COE}}$ is the center-of-energy-deposited distance from the beam, based on the photon position at the calorimeter ($x_{\gamma_{1,2}}$, $y_{\gamma_{1,2}}$) and the final photon energies, $E_{\gamma_{1,2}}$. The photon separation angle projection on the calorimeter plane and the angle between the beam axis and the $\pi\nu\nu$-hypothesis-reconstructed photon momenta are denoted as $\theta_{\gamma,{\rm calo}}$ and $\theta_{\gamma,{\rm beam}}$, respectively. The $z_{\text{vtx}}$ position is calculated assuming that we are reconstructing an on-axis $\pi^0 \to \gamma\gamma$ decay, where $z_{\text{vtx}} = 0$ corresponds to $z = 21\,\m$ from the T1 target for Step 1 and $44\,\m$ for Step 2. Finally, we quote $\mathcal{A}_\text{add}$, the selection efficiency of additional shape-related cuts (cluster shape, pulse shape and shower depth). Since we do not have more detailed information about $\mathcal{A}_\text{add}$, we assume a uniform distribution over the whole signal region, using the numbers quoted in Refs.~\cite{KOTO:2018dsc,Aoki:2021cqa}. As we can see from Table~\ref{tab:KaonModesSelections}, the main difference between the current Step 1 data and the future planned run is the collected statistics in terms of the number of protons on target $\NPoT$ and the selection efficiency of the shape-related cut algorithms which has improved considerably while keeping a good rejection power for hadronic backgrounds. 

In the search for $a\to \gamma\gamma$ signal events, we are not limited to the $K_L \to \pi^0 \nu\bar{\nu}$ signal region, since the $a\to \gamma\gamma$ events have different kinematics; for details, see Sec.~\ref{sec:flux}. Therefore, we need to estimate the expected $N_{\rm exp}$ and the observed $N_{\rm obs}$ number of SM events in the whole $z_{\text{vtx}}$-$p_{T,\pi^0}$ plane for the various KOTO datasets. For the Step 1 2015 dataset, we make estimations in all regions in the $z_{\text{vtx}}$-$p_{T,\pi^0}$ plane, using the $N_{\rm exp}$ and $N_{\rm obs}$ shown in Fig.~3 of Ref.~\cite{KOTO:2018dsc}. In addition, we estimate the sensitivity of the region of $p_{T,\pi^0} > 0.5\,\GeV$ that is out of the range of the referential figure. Since the pion transverse momentum is expected to be smaller than $0.5\,\GeV$ for most of the known physics processes~\cite{NA62KLEVER:2022nea}, we assume that there is no SM background in this region i.e., $N_{\rm exp} = 0$.

In order to project the KOTO sensitivity for the future Step 1 dataset, we use the $N_{\rm exp}$ backgrounds in the various $z_{\text{vtx}}$-$p_{T,\pi^0}$ regions which were presented in Ref.~\cite{KOTO:2020prk} for the $N_{\text{PoT}} = 3.05 \times 10^{19}$ statistics and rescale these numbers to the expected statistics $N_{\text{PoT}} = 14 \times 10^{19}$~\cite{KOTO:ichep22}. If we exclude $\pi\nu\bar{\nu}$ and the surrounding region ($z_{\text{vtx},\pi\nu\bar{\nu}} < 5.1\,\m$, and $p_{T,\pi^0} < 0.26\,\GeV$), where the number of background events is large compared to the expected number of $a \to \gamma\gamma$ events,\footnote{The SM $K_L \to \pi^0 \nu\bar{\nu}$ is considered to be a background for the search for $a \to \gamma\gamma$ decay.} we get $N_{\rm exp} = 3.35$ events with $N_{\text{PoT}} = 14 \times 10^{19}$.

For KOTO Step 2, we use Ref.~\cite{NA62KLEVER:2022nea} to estimate the background rate in the $1.75\;\mathrm{m} < z_{\text{vtx}} < 15\,\m$ and $p_{T,\pi^0} > 0.4\,\GeV$ regions. We find $N_{\rm exp}\approx 1.38$ events with $N_{\text{PoT}} = 6 \times 10^{20}$ statistics (assuming again that there is no background in the $p_{T,\pi^0} > 0.5\,\GeV$ region for $z_{\text{vtx}} < 15\,\m$).

%%%%%%%%%%%%%%%%%%%%%%%%%%%%%%%%%%%%%%%%%%%%%%%%
\begin{table}[t]
    \centering
    \begin{tabular}{ccc}
         \hline\hline
         & Step 1 & Step 2 \\
         \hline\hline
         $\sqrt{x_{\gamma_{1,2}}^2+y_{\gamma_{1,2}}^2}$ & $<0.85\,\m$ & $<1.35\,\m$ \\
         $\min(\abs{x_{\gamma_{1,2}}},\abs{y_{\gamma_{1,2}}})$ & $>0.15\,\m$ & $>0.175\,\m$ \\ 
         $R_{\text{COE}}$ & $>0.2\,\m$ & $-$ \\
         $r$ & $>0.3\,\m$ & $>0.3\,\m$ \\
         $\theta_{\gamma,{\rm calo}}$ & $<150^{\circ}$ & $<150^{\circ}$ \\
         $E_{\gamma_1} + E_{\gamma_2}$ & $> 0.65\,\GeV$ & $> 0.5\,\GeV$\\
         $E_{\gamma_{1,2}}$ & $\in[0.1,2.0]\,\GeV$ & $>0.1\,\GeV$ \\
         $E_{\gamma_1}/E_{\gamma_2}$ & $>0.2$ & $-$ \\
         $z_{\text{vtx}}$ & $\in[2.9,6.0]\,\m$ & $\in[1.75,15]\,\m$ \\
         $E_\gamma\theta_{\gamma,{\rm beam}}$ & $> 2.5^{\circ}$~GeV & $-$ \\
         $\mathcal{A}_\text{add}$ & 0.52\,(0.9) & 0.73 \\ 
         \hline
         $\NPoT \times 10^{19}$ & $2.2\,(14)$ & $60$ \\
         \hline\hline
    \end{tabular}
    \makeatletter\long\def\@ifdim#1#2#3{#2}\makeatother
    \caption{The selection conditions for $K_L \to \pi^0 \nu\bar{\nu}$ for Step 1~\cite{KOTO:2018dsc} and Step 2~\cite{Aoki:2021cqa}. 
    The $\NPoT$ and shape-related cut efficiency (in parenthesis) are for the future Step 1 run~\cite{KOTO:ichep22}. }
    \label{tab:KaonModesSelections}
\end{table}
%%%%%%%%%%%%%%%%%%%%%%%%%%%%%%%%%%%%%%%%%%%%%%%%

\begin{figure*}[t]
\begin{center}
\includegraphics[width=0.45\textwidth]{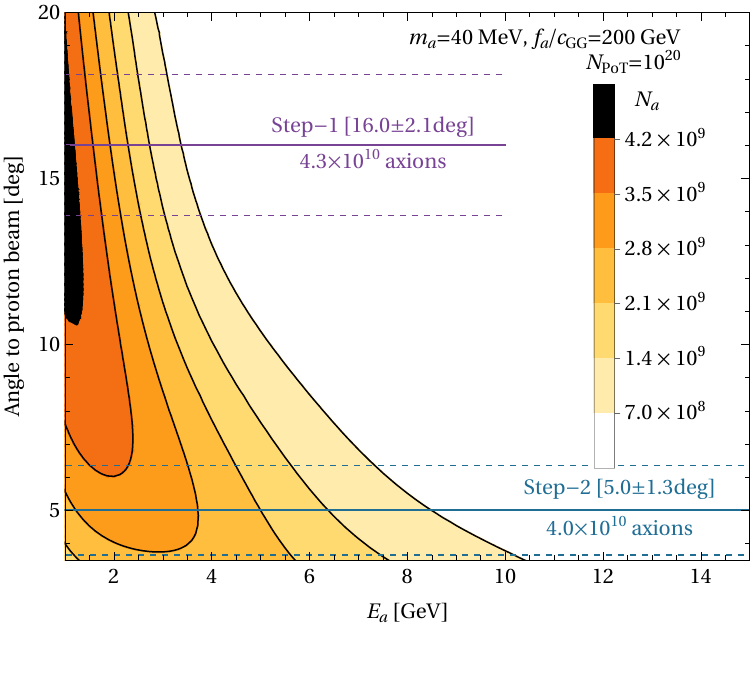}
\includegraphics[width=0.45\textwidth]{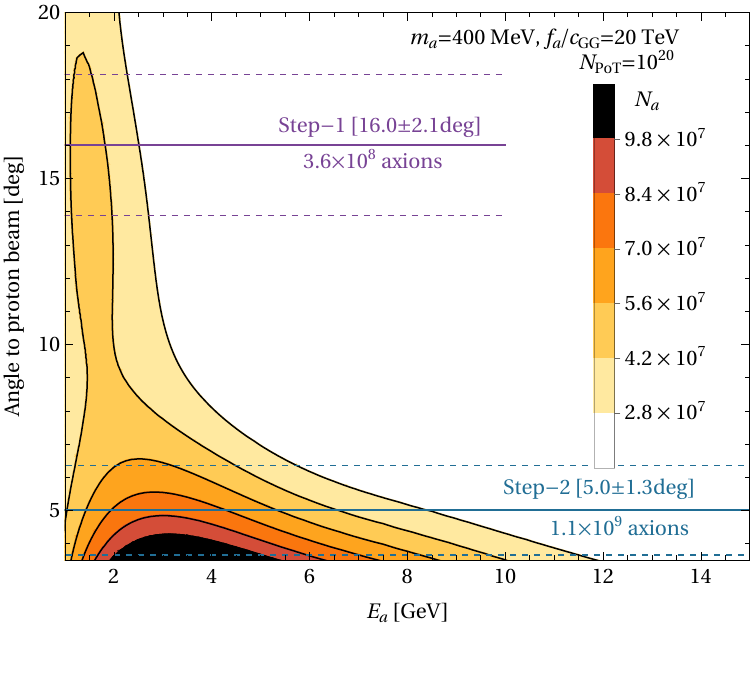}
\end{center}
\vspace{-10mm}
\caption{\label{fig:angleEa}
Expected axion production yield $N_a$ at the target as a function of axion energy, $E_a$, and its angle to the incident proton beam for two cases of axion masses and couplings. The respective angles at which the detector is placed in Step 1 and Step 2 are shown by solid lines with dashed lines indicating the part of the distribution of interest given the ECAL edges.}
\vspace{-2mm}
\end{figure*}

%%%%%%%%%%%%%%%%%%%%%%%%%%%%%%%%%%%%%%%%%%%%%%%%
\subsubsection{Beam-dump mode}
\label{sec:bd-mode}
%%%%%%%%%%%%%%%%%%%%%%%%%%%%%%%%%%%%%%%%%%%%%%%%

In the beam dump mode, so far KOTO has collected data corresponding to $N_{\text{PoT}} = 2.2 \times 10^{17}$~\cite{KOTO:beamdump}, which is about 2 orders of magnitude less than in the kaon mode. For our projection of the beam dump mode, we assume that 10 times more data will be collected in this mode (i.e. $N_{\text{PoT}} = 2.2 \times 10^{18}$), while keeping the background under control, i.e. we consider a background-free search. We explore this case for both KOTO Step 1 and KOTO Step 2 layouts,\footnote{The choice of $\mathcal{A}_\text{add}$ is because we do not find enough information in Ref. \cite{KOTO:beamdump} to derive it. Also, we later show that even with this optimistic choice of $\mathcal{A}_\text{add}$, the sensitivity at the dump mode is not competitive with that of the kaon mode.
}  assuming $\mathcal{A}_\text{add} = 1$.
The selection conditions are simply both photons being in the calorimeter acceptance with cluster distance $> 0.3\,\m$ (as used in the $\pi^0 \nu \bar{\nu}$ analysis). For KOTO Step 1 dump-mode we require at least $50\,\MeV$ deposited on the calorimeter per photon; and for KOTO Step 2, we require at least $100\,\MeV$ per photon and at least $500\,\MeV$ in total deposited on the calorimeter.

%%%%%%%%%%%%%%%%%%%%%%%%%%%%%%%%%%%%%%%%%%%%%%%%
\section{Axion production and detection}
\label{sec:flux}
%%%%%%%%%%%%%%%%%%%%%%%%%%%%%%%%%%%%%%%%%%%%%%%%

The conventional axion production at the $K_L$ experiments is from $K_L\to \pi^0 a$ decay, which is $CP$-violating and thereby suppressed. 
Here, the relevant production to probe the long-lived axion occurs at the fixed target T1 where $K_L$ is produced i.e., in the proton-gold collisions. We consider two production mechanisms: Primakoff production and axion-meson mixing, both are implemented in the {\sc Alpinist} framework, which uses {\sc Pythia~8}~\cite{Sjostrand:2014zea} to generate meson distributions.
In the following, we show the axion production yields for these mechanisms for the $30\,\GeV$ proton beam. The validation of obtained yields with the detector under the $16^\circ$ angle is given in Appendix~\ref{sec:app_validation}. As shown in Ref.~\cite{Dobrich:2019dxc}, the yield and the momentum spectrum is well described by {\sc Pythia~8} also for angles smaller than $16^\circ$ and higher beam energy. 

%%%%%%%%%%%%%%%%%%%%%%%%%%%%%%%%%%%%%%%%%%%%%%%%
\subsection{Axion production in the target}
%%%%%%%%%%%%%%%%%%%%%%%%%%%%%%%%%%%%%%%%%%%%%%%%

The gluon coupling $c_{GG}$ induces the axion mixing to the neutral mesons of the same quantum numbers, $P \in \left\lbrace \pi^0, \eta, \eta' \right\rbrace$. The axion yield from mixing production is then approximately given by
\begin{align}
    N_{a}^{\text{mix}} 
    \approx
    N_{\pi^0}\cdot|\theta_{a\pi}|^2 +  N_{\eta}\cdot|\theta_{a\eta}|^2 + N_{\eta^{\prime}}\cdot|\theta_{a\eta^{\prime}}|^2
\end{align}
where $N_{\pi^0}$ is the production yield of neutral pion, $\theta_{a\pi}$ is the pion-axion mixing angle, and similar notations are applied to $\eta$ and $\eta'$. To leading order in $f_\pi/f_a$, the mixing angles $\theta_{aP}$ are 
\begin{align}
    \theta_{aP} 
    \approx
   \frac{f_\pi}{f_a}\frac{K_{aP} m_a^2 + m_{aP}^2}{m_a^2-m_P^2},
\end{align}
where $f_{\pi}\approx 93\,\MeV$, and we use the same notation for the kinetic and mass mixing ($K_{aP}$ and $m_{aP}$) as in Refs.~\cite{Aloni:2018vki,Jerhot:2022chi}. The $\eta$--$\eta^\prime$ mixing angle used is $\sin\theta_{\eta\eta^\prime} = -1/3$. The kinematics for the axion mixing production is treated according to Appendix~D of Ref.~\cite{Jerhot:2022chi}.
The Primakoff process is governed by the axion-photon interaction
\begin{align}
    c_{\gamma\gamma}\frac{\alpha_{\rm EM} a}{8\pi f_a}F_{\mu\nu}\tilde F^{\mu\nu},
\end{align}
where $\alpha_{\rm EM}$ is the fine-structure constant and $c_{\gamma\gamma}$ is the effective axion-photon coupling. The axion-photon coupling is generated by $c_{BB}$ and $c_{WW}$ couplings and for $m_a\lesssim m_\rho$ also by the low-energy contribution of $c_{GG}$~\cite{Ertas:2020xcc,Bauer:2017ris,Domingo:2016yih}:
\begin{align}
    c_{\gamma\gamma} 
=&   c_{BB} + c_{WW}
    -c_{GG}\left(1.92 + 2\sum_P 
    \frac{f_a}{f_P} \theta_{aP}\right),
  \label{eq:cgamma}
\end{align}
where $f_{\eta}\approx f_\pi$  and $f_{\eta^{\prime}}\approx 73\,\MeV$. There are two sources of photons in the target that can produce axions in the interaction with the target nuclei via the Primakoff process: off-shell photons from the proton of the primary beam~\cite{Dobrich:2015jyk} and photons from decays of secondary neutral pseudoscalars produced in the target. The latter has been shown to be dominating~\cite{Dobrich:2019dxc}. The axion yield from these processes is given by their sum and is denoted as $N_a^\text{Prim}$.

Additional production could be from flavor-changing kaon decays near the fixed target. With the gluon coupling $c_{GG}$, the $K^\pm(K_S) \to \pi^\pm (\pi^0) a$ is  not suppressed by a loop or $CP$ violation~\cite{Georgi:1986df,Bauer:2021wjo,Goudzovski:2022vbt}. The $c_{WW}$ coupling induces the same processes at one loop~\cite{Izaguirre:2016dfi}. These production rates could be sizable because the total width of kaons is small; i.e., the BR is enhanced. However, most kaons are removed by the collimators or deflected by the magnetic field. Including these effects, we estimate that $K^+\to \pi^+ a$ is subdominant. Still, the production from $K_S$ is potentially interesting due to the shorter lifetime, but it requires a simulation of $K_S$ transport, which is beyond the scope of this paper. Therefore,  we neglect axion production from $K^+$ and $K_S$ decays. 

To summarize, the total axion yield at the T1 target is approximated to be
\begin{align}
    N_a \approx N_a^\text{mix} + N_a^\text{Prim}\,, 
\end{align}
where the exact relative contributions of the two mechanisms depend on the specific model. We show the differential distributions with respect to the axion energy ($E_a$) and its production angle to the proton beam ($\theta_a$, see Fig.~\ref{fig:illustration}) in Fig.~\ref{fig:angleEa} for two benchmarks with axion masses of 40\,MeV and 400\,MeV and the couplings being dominated by the gluon, $f_a/c_{GG}=200\,{\rm GeV}$ and $20\,{\rm TeV}$, respectively. In this coupling benchmark for the $m_a=400$\,MeV case, the overall production yield from the mixing with $\pi^0$ and $\eta$ exceeds significantly the Primakoff production while for smaller axion masses, the Primakoff production becomes more relevant, similarly to what has been observed in Ref.~\cite{Jerhot:2022chi} for experiments operating with higher beam energies.

%%%%%%%%%%%%%%%%%%%%%%%%%%%%%%%%%%%%%%%%%%%%%%%%
\begin{figure}[t]
\begin{center}
\includegraphics[width=0.45\textwidth]{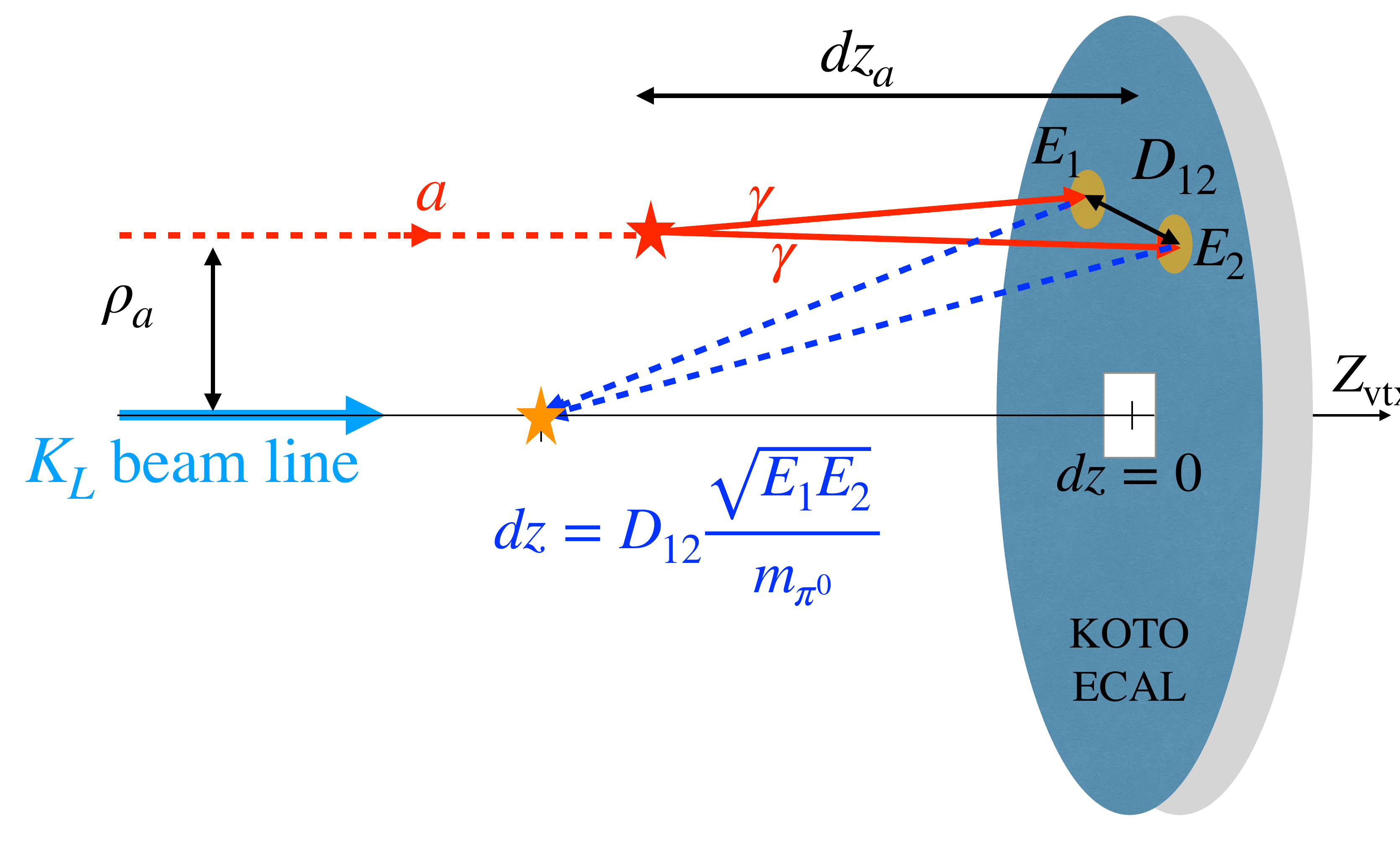}
\vspace{-2mm}
\caption{\label{fig:schematic} A schematic picture of the axion event in the kaon mode. The disk represents the ECAL, and the axis is the $K_L$ beamline. The distance from the ECAL to the upstream is $dz$. A typical axion trajectory is almost parallel to the beam axis in the distance $\rho_a$ (see Fig.~\ref{fig:illustration}). If the final-state photons from the axion decay at $dz_a$ leave the energy of $E_{1,2}$ with a separation of $D_{12}$ on the ECAL, the vertex position $dz$ is reconstructed on the beam axis.}
\vspace{-2mm}
\end{center}
\end{figure}
%%%%%%%%%%%%%%%%%%%%%%%%%%%%%%%%%%%%%%%%%%%%%%%%

%%%%%%%%%%%%%%%%%%%%%%%%%%%%%%%%%%%%%%%%%%%%%%%%
\subsection{Axion detection mechanism}
%%%%%%%%%%%%%%%%%%%%%%%%%%%%%%%%%%%%%%%%%%%%%%%%

Given the beam energy and the distance between the axion production location and the detector, the KOTO experiment can search for long-lived axions. Below, we see that for the relevant masses, the dominant decay channel is $a\to\gamma\gamma$, which is given by
\begin{align}
    \Gamma_{\gamma\gamma}=\frac{\alpha_{\rm EM}^2m_a^3}{256\pi^3}\frac{c_{\gamma\gamma}^2}{f_a^2}\,,
\end{align}
where the effective photon coupling $c_{\gamma\gamma}$ is defined in Eq.~\eqref{eq:cgamma}. For an axion heavier than 1\,GeV with nonzero $c_{GG}$ coupling, the width of hadronic decay modes, such as $a\to \pi \pi \eta$, estimated in Ref.~\cite{Aloni:2018vki}, dominates the total width. However, the relevant final state for search at KOTO is diphoton, which typically becomes a subdominant mode for $m_a\gtrsim0.5\,\GeV$ (with large model dependence), resulting in reduced sensitivity. Details on KOTO sensitivity for hadronic axion decays can be found in Appendix~\ref{sec:app_hadrons}.
 
Long-lived axions produced at the fixed target can reach the distant decay volume, and a diphoton decay leaves a characteristic signal, see a schematic picture of the axion decay event in Fig.~\ref{fig:schematic}. When axions enter the decay volume, they are almost parallel to the $K_L$ beam axis, because the distance to the ECAL is larger than the ECAL size, but away from the axis with the distance $\rho_a$. The two photons from an $a \to \gamma\gamma$ decay in the decay volume can then hit the ECAL, mimicking the signal of $K_L\to \pi^0\nu\bar\nu$. If the standard reconstruction algorithm for $K_L\to \pi^0\nu\bar\nu$ is applied, the reconstructed position ($dz$) will be different from the true distance between the axion decay point and the ECAL ($dz_a$), but the event is not discarded. In this sense, this signal is similar to the halo $K_L\to \gamma\gamma$ background that the KOTO Collaboration found in the earlier data, but the reconstructed distributions are typically different. See Fig.~29 of Ref.~\cite{NA62KLEVER:2022nea} for the distribution. 

Contrary to the kaon mode, the detection mechanism in the beam dump mode is relatively straightforward, as it is a dedicated run to search for long-lived particles decaying to photons. 

%%%%%%%%%%%%%%%%%%%%%%%%%%%%%%%%%%%%%%%%%%%%%%%%
\begin{figure*}[t]
\begin{center}
\includegraphics[width=0.45\textwidth]{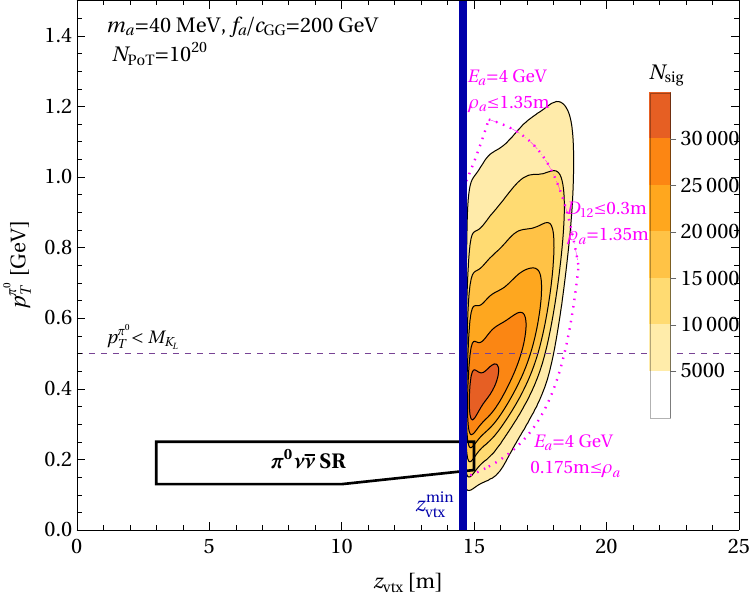}
%\vspace{5pt}
\includegraphics[width=0.45\textwidth]{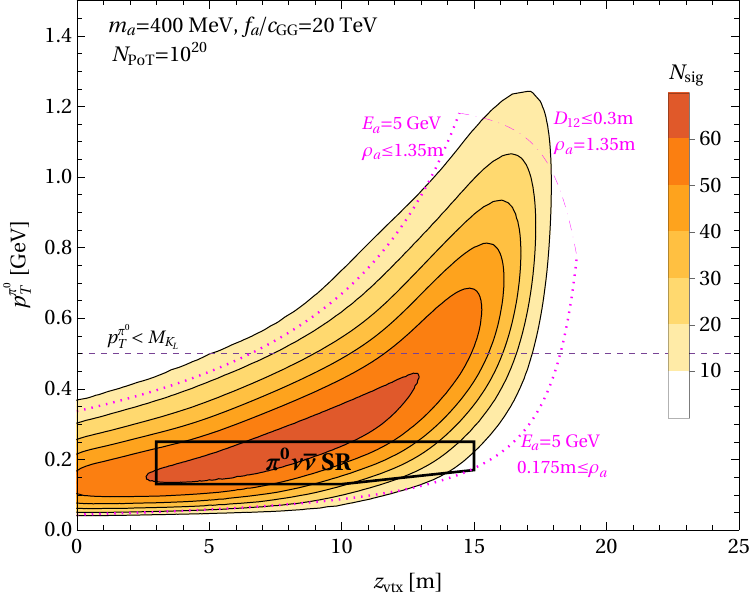}
\vspace{-2mm}
\caption{\label{fig:pTZ} Axion signal event distributions $N_\text{sig}$ as could be found in the $\pi\nu\bar{\nu}$ analysis at KOTO Step 2 for specific axion models. For the analytic understanding of the distributions, we show the lines corresponding to Eq.~\eqref{eq:pTvsZ} with $\rho_a =0.175,1.35$\;m and a specific $E_a$ as dashed magenta lines.}
\vspace{-2mm}
\end{center}
\end{figure*}
%%%%%%%%%%%%%%%%%%%%%%%%%%%%%%%%%%%%%%%%%%%%%%%%

%%%%%%%%%%%%%%%%%%%%%%%%%%%%%%%%%%%%%%%%%%%%%%%%
\subsection{Signal from axion decay in kaon mode}
%%%%%%%%%%%%%%%%%%%%%%%%%%%%%%%%%%%%%%%%%%%%%%%%

The long-lived axion $a \to \gamma\gamma$ decay passes the event selections described in Sec.~\ref{sec:Kaon-mode} when $\rho_a>0$ which introduces factitious transverse momentum of the diphoton system. Some selection criteria are universal, often limited by the experimental resolution, but the remaining cuts assume the topology of $K_L\to \pi^0(\to \gamma\gamma)\nu\bar\nu$. The KOTO experiment could implement a dedicated analysis for the long-lived particles, but here we simply adopt the standard $K_L\to \pi^0\nu\bar\nu$ analysis because the backgrounds are well investigated. 
Assuming $K_L\to \pi^0\nu\bar\nu$ topology for the long-lived axion decays leads to several nontrivial characteristics of the axion signal event distribution $N_\text{sig}$ in the $z_{\text{vtx}}$-$p_{T,\pi^0}$ plane. 
Example distributions for parameters for which we expect interesting sensitivities of Step 2 are shown in Fig.~\ref{fig:pTZ}. In the following, we give analytic understandings of the characteristics based on several simplifications. 

We assume that the axion enters the decay volume in parallel to the beamline with distance $\rho_a$. The axion invariant mass is given by the photon energies, $E_{1,2}$, and the opening angle, $\theta_{12}$, 
\begin{align}
    m_a^2 = 2E_1E_2(1-\cos\theta_{12}) \simeq E_1E_2\theta_{12}^2\, .
\end{align}
Using this, the separation of the two photons on the ECAL, $D_{12}$, is approximately
\begin{align}
    \label{eq:D12}
    D_{12} \simeq dz_a  \theta_{12} \simeq dz_a \frac{m_a}{\sqrt{E_1E_2}} \, .
\end{align}
Then, the {\it reconstructed} vertex position, assuming the event topology of $K_L\to \pi^0\nu\bar\nu$, is given by 
\begin{align}
    \label{eq:dzdza}
    dz \equiv D_{12} \frac{\sqrt {E_1E_2}}{m_{\pi^0}} \simeq dz_a \frac{m_a}{m_{\pi^0}} \, .  
\end{align}
The distance from the ECAL ($dz$) is translated to  the standard coordinate system by $z_{\text{vtx}}=L-dz$ where $L=$6\;(20)\;m in Step 1 (Step 2). Because only the axion decays in the decay volume are accepted, there exists a limitation of $dz_a< dz_a^{\rm max}=2.9\,(18.25)\,\m$ in Step 1 (Step 2). This limitation, together with Eq.~\eqref{eq:dzdza}, gives us a condition on the maximum spread of the distribution over the reconstructed vertex position as $z_{\text{vtx}}> L-dz_a^{\rm max}({m_a}/{m_{\pi^0}})$. The boundary $z_{\text{vtx}}^\text{min}$ is shown as a blue vertical line in the left panel of Fig.~\ref{fig:pTZ}. 

Another feature can be seen as a correlation with both $p_{T}^{\pi^0}$ and $z_{\text{vtx}}$. In the case of $K_L\to \pi^0\nu\bar\nu$, the transverse kick is from $K_L$ decay, and hence, $p_T^{\pi^0}<m_{K_L}$. However, the transverse asymmetry of the ECAL hits is merely from the transverse position of the incident axion. Supposing the distance from the axion to the beamline is $\rho_a$, the reconstructed $p_T^{\pi^0}$ is roughly 
\begin{align}
    p_T^{\pi^0} 
    \simeq  \frac{E_a \rho_a}{\sqrt{\rho_a^2 +dz^2}}
    \simeq \frac{E_a}{\sqrt{1+(L-z_{\text{vtx}})^2/\rho_a^2}}.  \label{eq:pTvsZ}
\end{align}
where $E_a=E_1+E_2$ is the axion energy. The correlation between $p_T^{\pi^0}$ and $z_{\text{vtx}}$ is explained by the above formula. Note that $p_T^{\pi^0}$ can easily exceed $m_{K_L}$, because $E_a\sim \cO(1)\,\GeV$.

%%%%%%%%%%%%%%%%%%%%%%%%%%%%%%%%%%%%%%%%%%%%%%%%
\begin{figure*}[t]
\begin{center}
\includegraphics[width=0.4\textwidth]{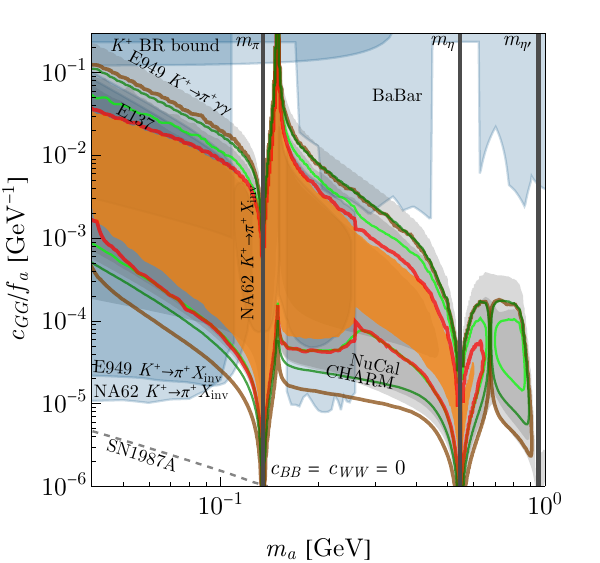}
\includegraphics[width=0.4\textwidth]{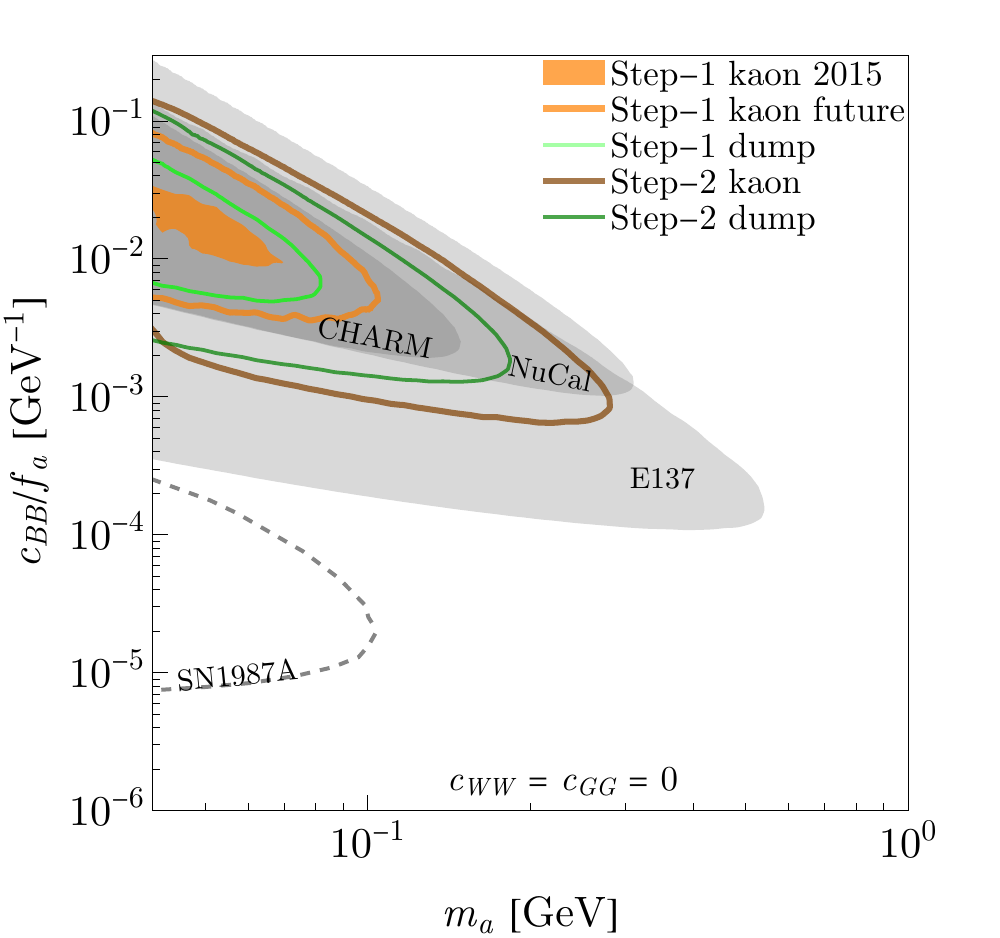}
\includegraphics[width=0.4\textwidth]{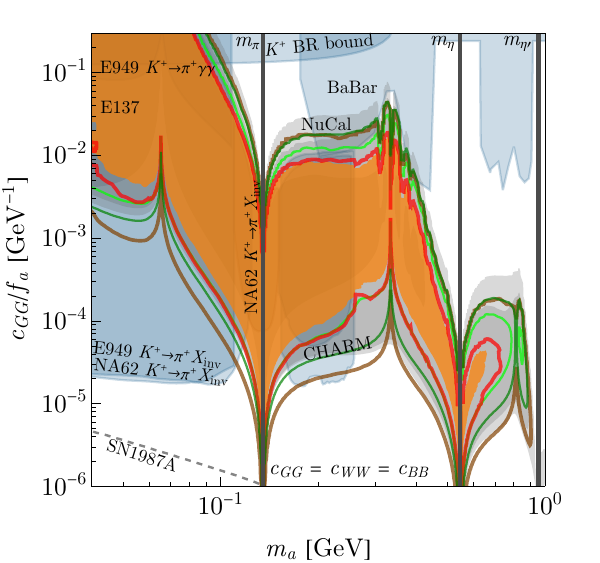}
\vspace{-1mm}
\caption{\label{fig:excl_BB_GG_COD}90\% C.L. exclusion bounds (filled contours) 
and projected limits (empty contours) for scenarios (i)-(iii) for all KOTO setups considered compared to the exclusions from beam dump experiments, exclusion from $B\to K a$ and $K \to \pi a$ decays and the bound from the supernova SN1987A (shown as a dashed line as it is affected by significant uncertainties see, \eg, Refs.~\cite{Ertas:2020xcc,Bar:2019ifz}).}
\vspace{-5mm}
\end{center}
\end{figure*}
%%%%%%%%%%%%%%%%%%%%%%%%%%%%%%%%%%%%%%%%%%%%%%%%

The last feature that can be observed is the vanishing of the distribution at large $p_T^{\pi^0}$ and $z_{\text{vtx}}$, which is a consequence of the two-photon separation cut, $D_{12}\leq0.3$\,m. This can be understood by combining Eqs.~\eqref{eq:D12},~\eqref{eq:dzdza}, and~\eqref{eq:pTvsZ} with a simplification of $E_{1,2}\approx E_a/2$ as
\begin{align}
	\label{eq:D12cut}
	p_T^{\pi^0}\lesssim \frac{2m_{\pi^0}}{0.3\;\mathrm{m}\sqrt{(L-z_{\text{vtx}})^{-2}+\rho_a^{-2}}} \, .
\end{align}
Therefore, the distribution between the two dashed lines from Eq.~\eqref{eq:pTvsZ} vanishes at large $p_T^{\pi^0}$ and $z_{\text{vtx}}$ at an approximate bound corresponding to Eq.~\eqref{eq:D12cut} with $\rho_a=1.35\,\m$. 

Since $p_T^{\pi^0}$ of the long-lived axion events can be significantly larger compared to the $K_L$ events or background, we use the high $p_T^{\pi^0}$ region. 
The detail of the signal region and the expected background yield is discussed in Sec.~\ref{sec:Kaon-mode}.

%%%%%%%%%%%

%%%%%%%%%%%%%%%%%%%%%%%%%%%%%%%%%%%%%%%%%%%%%%%%
\section{Bounds and projections for axions} 
\label{sec:limits}
%%%%%%%%%%%%%%%%%%%%%%%%%%%%%%%%%%%%%%%%%%%%%%%%

\begin{figure*}[t]
\begin{center}
\includegraphics[width=0.4\textwidth]{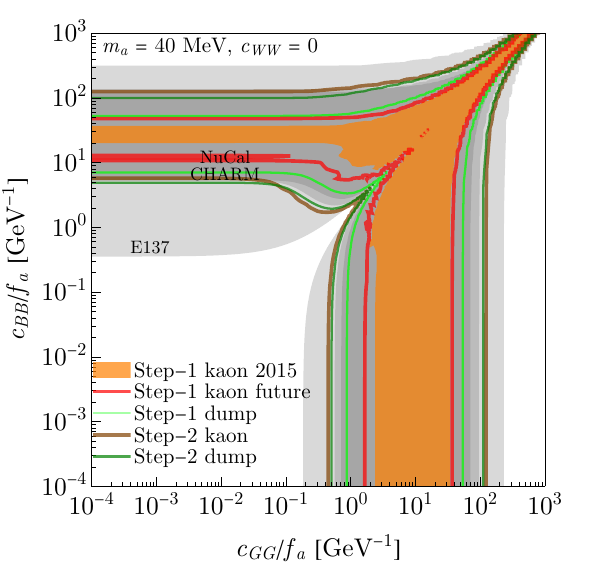}
\includegraphics[width=0.4\textwidth]{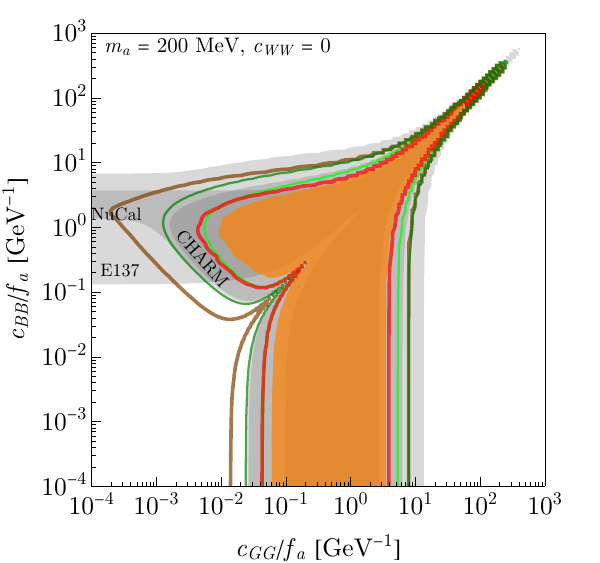}
\includegraphics[width=0.4\textwidth]{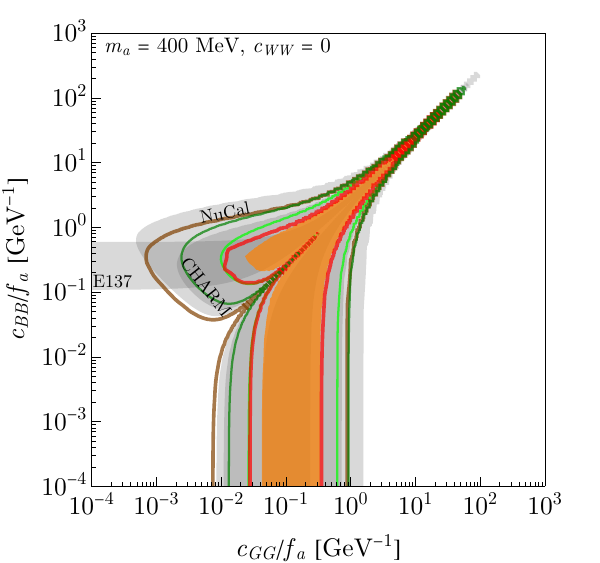}
\vspace{-1mm}
\caption{\label{fig:excl_fixed_ma} 90\% C.L. exclusion bounds and projected limits for fixed axion mass and variable $c_{BB}$ and $c_{GG}$ couplings. }
\vspace{-5mm}
\end{center}
\end{figure*}

We derive the bounds for different axion models using the current KOTO Step 1 results and project the sensitivities of the future runs. For this purpose, we simulate the signal, including all production processes i.e., Primakoff and meson mixing mechanisms. The various experimental setups are implemented in the {\sc Alpinist} framework where we have also generated the datasets\footnote{For increasing the precision of the estimated number of observed events, each mass and decay width bin is evaluated with $2\times 10^6$ axion decay events.} with the number of events expected $N_\text{sig}$ for each setup and production mode. These tables can be further used for showing the $N_\text{sig}$ distributions for any setup of model-dependent parameters using the rescaling module of the framework. In Fig.~\ref{fig:excl_BB_GG_COD}, we show the current bounds and projections for the expected sensitivity with future data corresponding to the following three benchmark models:
\begin{enumerate}
[label=(\roman*)]
 \item Heavy QCD axion: $c_{GG}\neq 0$ with $c_{BB} = c_{WW} = 0$.
 \item Hypercharge dominant: $c_{BB}\neq0$ with $c_{WW}=c_{GG}=0$.
  \item Codominance: $c_{BB} = c_{WW} = c_{GG}$.
\end{enumerate}
In Fig.~\ref{fig:excl_fixed_ma}, we show the current and projected results for several fixed masses for variable $c_{GG}$ vs $c_{BB}$ couplings, which is expected to be similar to the case of $c_{GG}$ vs $c_{WW}$, up to the FCNC production.

We compare our KOTO bounds and projections to the existing bounds from different experiments. We consider electron beam dumps E137~\cite{Bjorken:1988as} and E141~\cite{Riordan:1987aw}, where we use the {\sc Alpinist} framework for the interpretation of the data provided in Ref.~\cite{Dolan:2017osp}, implementation that has been already done for Ref.~\cite{HIKE:2022qra}, and we perform a dedicated analysis to derive the proton beam dump bounds based on CHARM~\cite{CHARM:1985anb} and NuCal~\cite{Blumlein:1990ay}, as done in~ Ref.~\cite{Jerhot:2022chi} (bounds in gray shade). 

The axion can be produced by flavor-changing meson decays, especially in the presence of $c_{WW}$ and $c_{GG}$ (the corresponding bounds are shown in the blue shade). We adopt the results of $K^+\to \pi^+ a$ from Ref.~\cite{Bauer:2021wjo} and apply the bounds of E949~\cite{BNL-E949:2009dza, E949:2005qiy} and NA62~\cite{NA62:2020pwi, NA62:2021zjw}. The scheme of the recast is found in Ref.~\cite{Gori:2020xvq}. For $B\to K a(\to \gamma\gamma)$, the \textit{BABAR} bound can be used~\cite{BaBar:2021ich, Bertholet:2021hjl}. The two-loop production calculation with $c_{GG}$ is found in Ref.\cite{Chakraborty:2021wda}. The contribution from $c_{WW}$ at one loop is calculated in~\cite{Izaguirre:2016dfi}, but it is numerically subdominant for the benchmark~(iii), so for simplicity, we approximate $B\to Ka$ by the $c_{GG}$ contribution. The total width $K^+$ would be modified significantly for low $f_a/c_{GG}$, which leads to a relevant bound at $m_a\sim m_{\pi^0}$. Requiring ${\rm BR}(K^+\to \pi^+ a)<3\times 10^{-3}$, based on Sec.~2.2.2 of Ref.~\cite{Goudzovski:2022vbt} results in the bound $f_a/c_{GG}\lesssim 5$~GeV. The $c_{BB}$-only scenario is not significantly constrained by the meson decays, since the production originates from electroweak two-loop diagrams. Therefore, in Fig.~\ref{fig:excl_fixed_ma}, the meson decay bounds are omitted because the corresponding bounds in the limit of the benchmark~(ii) are unknown. Finally, the shown SN1987A bounds are those derived in Ref.~\cite{Ertas:2020xcc} but are plotted with a dashed line as their robustness is under debate~\cite{Bar:2019ifz}.

We find that at Step 1, KOTO cannot probe new regions in the parameter space, and it is sensitive only in regions already covered by other proton beam dump experiments for all coupling scenarios. The \textit{non-}observation of additional signal on top of the expected backgrounds in the past KOTO $K_L \to \pi^0\bar{\nu}\nu$ analyses only confirms the results of these past experiments. While for scenarios with photon coupling domination in the future, KOTO Step 2 cannot compete with the past electron beam dump experiments E137 and E141 either, it can probe new regions of parameter space for larger masses ($m_a \gtrsim m_{\pi^0}$) for scenarios with gluonic coupling thanks to enhanced axion production through mixing with other neutral pseudoscalars.

%%%%%%%%%%%%%%%%%%%%%%%%%%%%%%%%%%%%%%%%%%%%%%%%

%%%%%%%%%%%%%%%%%%%%%%%%%%%%%%%%%%%%%%%%%%%%%%%%

%%%%%%%%%%%%%%%%%%%%%%%%%%%%%%%%%%%%%%%%%%%%%%%%
\section{Conclusions}
\label{sec:conclusion}
%%%%%%%%%%%%%%%%%%%%%%%%%%%%%%%%%%%%%%%%%%%%%%%%

In this paper, we show that the KOTO experiment, beyond its conventional purposes, can perform long-lived particle searches in its two different data-taking modes, the kaon and the beam dump. In both modes, NP particles are produced at the proton target interaction point and can decay in the whole decay volume of the detector (see Figs.~\ref{fig:illustration} and~\ref{fig:schematic}). We show that the future KOTO runs will explore the uncharted parameter space of sub-GeV axions, which may address the strong $CP$ problem.

First, we show that the kaon mode, where the majority of the KOTO data are taken, is sensitive not only to $K_L\to\pi^0\nu\bar\nu$ and $K_L\to \pi^0 a$~\cite{Gori:2020xvq} but also to axions, which are originating from the interaction in the proton target and mimic the rare kaon decay signal. The main difference between the two signals is that the axion events extend the distribution of $p_T^{\pi^0}$ greater than $m_{K_L}$, as shown in Fig.~\ref{fig:pTZ}. This region is unphysical for the diphoton events from the kaon decays; thus, we assume no SM background there. Even though our derived constraints based on KOTO 2015 dataset are currently not competitive, they reaffirm the constraints obtained with experiments of very different topology and proton impact energy. We have also shown in Fig.~\ref{fig:excl_BB_GG_COD} that KOTO in Step 2 can indeed explore new parameter space for axions with $c_{GG}$ coupling and $m_a\gtrsim 100\,\MeV$, without changes of the main analysis steps.

Second, we have evaluated projections of KOTO running in the beam dump mode as recently presented by the collaboration~\cite{KOTO:beamdump}. Here we have shown that KOTO, due to its low proton beam energy and large angle between the beam and the detector, can especially well explore parameter space at very low couplings, complementary to such searches at higher energies \eg{}, NA62~\cite{Dobrich:2015jyk}, FASER~\cite{Feng:2018pew} or DarkQuest~\cite{Blinov:2021say}. Although the analysis in beam dump mode is suitable for searches for long-lived particles, the sensitivity to the axions is weaker than in the kaon mode because the expected statistics is significantly lower considering the KOTO physics goals.

In this study, we have focused on the axions to demonstrate the proof of concept, whereas similar analyses could be performed to update  the bounds of other long-lived particles from past proton-beam experiments. 
Furthermore, a dedicated analysis for long-lived particles rather than reinterpretation of the $K_L\to\pi^0\nu\bar\nu$ analysis could further improve the sensitivity although it requires additional background studies.
This work also updates the {\sc Alpinist} framework~\cite{zenodo} to include the KOTO geometry and the kinematics of the various processes. 
Thereby, other scenarios, including axions with different parameter combinations, can be easily studied. Following other patches to the {\sc Alpinist} code during the development, the sensitivity contours for other future and past experiments are also updated. All updates are publicly available in the framework repository~\cite{zenodo}.

\acknowledgments

Y.A. is supported by the National Science Foundation under Grant No. PHY-2013010. 
B.D. acknowledges funding through the European Research Council under Grant No. ERC-2018-StG-802836 (AxScale project) and the Lise Meitner program of the Max Planck society. 
J.J. acknowledges funding by the F.R.S.-FNRS, Belgium, through Grant No. FRIA/FC-36305.
Y.S. is supported by Grants from NSF-BSF~(No.~2021800), ISF~(No.~482/20), BSF~(No.~2020300), and the Azrieli foundation. 
K.T. acknowledges funding through  U.S. Department of Energy Grant No. DE-SC0010102 and Japan Society for the Promotion of Science KAKENHI Grant No. 21H01086.
J.J. and B.D. acknowledge useful discussions with Maksym Ovchynnikov and Tommaso Spadaro regarding updates of the ALPINIST code.
\bigskip
\appendix

\bigskip

\section{Validation of the Simulation}
\label{sec:app_validation}

\begin{figure}[b]
\begin{center}
\includegraphics[width=0.4\textwidth]{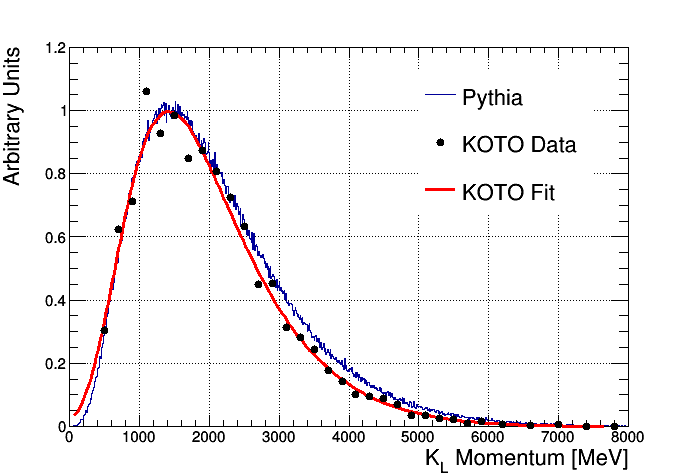}
\caption{\label{fig:validation} Comparison of the $K_L$ total momentum distribution from the simulation used in this work ({\sc Pythia~8}) and the data derived by KOTO~\cite{Masuda:2015eta} (KOTO Data). The red curve presents a fit of the data done by KOTO (KOTO Fit).}
\end{center}
\end{figure}

\begin{figure*}[t]
\begin{center}
\includegraphics[width=0.45\textwidth]{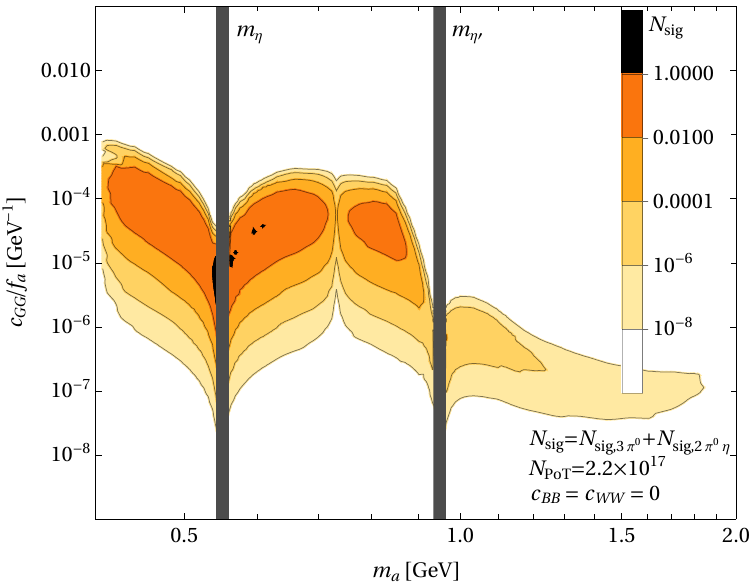}
\includegraphics[width=0.45\textwidth]{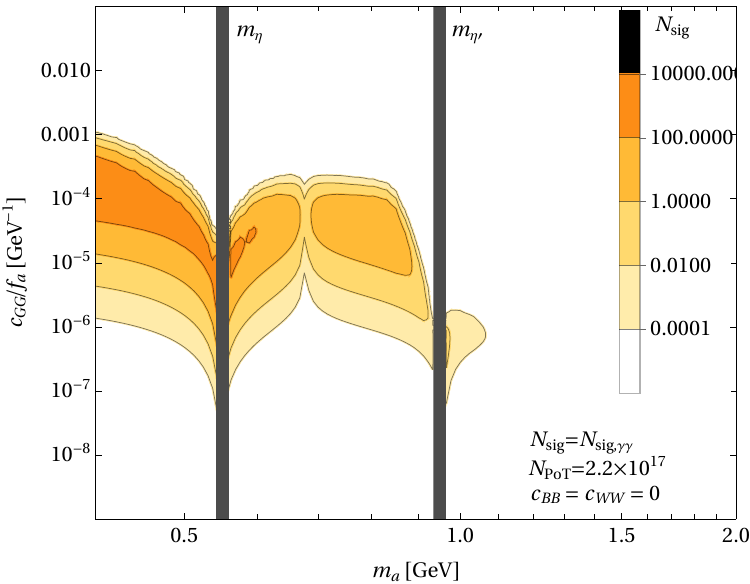}
\end{center}
\caption{\label{fig:sens} Expected number of signal events $N_\text{sig}$ for given values of axion mass $m_a$ and coupling $c_{GG}$ for scenario (i) with $N_\text{PoT} = 2.2 \times 10^{17}$ in KOTO Step 2 in beam dump mode. Comparison between $N_\text{sig}$ from $a\to \pi^0\pi^0\eta$ and $a\to 3\pi^0$ decays (left) and that from $a\to \gamma\gamma$ decay (right).
}
\end{figure*}

For the estimation of the axion flux from both the Primakoff and the mixing production, we first need to estimate the flux of the $\pi^0$, $\eta$ and $\eta^{\prime}$ mesons. For this purpose, we use {\sc Pythia~8}~\cite{Sjostrand:2014zea} for the generation of the $pp$ interactions. Since we cannot directly validate the meson flux simulation, we use the measured $K_L$ flux at KOTO~\cite{Masuda:2015eta} to normalize the meson yields. The measured number of $K_L$ collimated into $8\times 8\,\cm^2$~\cite{KOTO:2018dsc} at the end of a $20\,\m$ beamline is $(4.2 \pm 0.02_\text{stat} \pm 0.06_\text{sys}) \times 10^7$ $K_L$ per $2 \times 10^{14}$ protons on target, which corresponds to the measured number of $K_L$ per proton on target $N_{K_L}^\text{KOTO} \sim 2.1 \times 10^{-7}$. Therefore, we normalize the number of mesons simulated by {\sc Pythia~8} to the measured number as
\begin{align}
    N_P = \frac{N_{P}^\text{sim}}{N_{K_L}^\text{sim}} \times N_{K_L}^\text{KOTO},
\end{align}
where $N_{P}^\text{sim}$ is the number of ($\pi^0$, $\eta$, $\eta^{\prime}$) mesons produced per $pp$ interaction in the simulation. For $10^8$ simulated $pp$ interactions, after accounting for $K_L$ decays assuming the peak $K_L$ momentum $1.4\,{\rm GeV}$~\cite{KOTO:2018dsc} and $60\%$ loss of $K_L$ due to absorption in the beamline material~\cite{Masuda:2015eta}, we obtain the simulated number of $K_L$'s per $pp$ interaction, $N_{K_L}^\text{sim} \sim 2.0 \times 10^{-7}$. Therefore, $N_{K_L}^\text{KOTO}/N_{K_L}^\text{sim} \sim 1.05$, showing a good agreement between the total number of $K_L$'s measured and simulated using {\sc Pythia~8}. For the multiplicities of ($\pi^0$,~$\eta$,~$\eta^{\prime}$) mesons, based on the simulation, we obtain:
\begin{align}
    \frac{N_{\pi^0}^\text{sim}}{N_{K_L}^\text{sim}} \sim 21\,, \quad 
   \frac{N_{\eta}^\text{sim}}{N_{K_L}^\text{sim}} \sim 2.2\,, \quad 
   \frac{N_{\eta^{\prime}}^\text{sim}}{N_{K_L}^\text{sim}} \sim 0.17\,.
\end{align}
Furthermore, since axions interact rarely with ordinary matter, the surface of the axion flux potentially entering the detector $S_\text{axion}^\text{sim}$ is much larger, occupying the whole decay volume plane. When compared to the $K_L$ flux, which is collimated to the $S_{K_L}^\text{KOTO} = 8\times 8\,\cm^2$ profile at the end of the beamline, the ratio of the two surfaces is about $ S_\text{axion}^\text{sim} / S_{K_L}^\text{KOTO} \sim 490$.

Finally, in order to validate the kinematic distributions, we compare the distributions of $K_L$ obtained with {\sc Pythia~8} with the distributions measured by KOTO~\cite{Masuda:2015eta}. As shown in Fig.~\ref{fig:validation}, a good agreement is observed between the shapes of the distribution of the $K_L$ total momentum from the simulation used in this paper and the data measured by KOTO. This validation for $K_L$ gives some credibility also to the simulated ($\pi^0$,~$\eta$,~$\eta^{\prime}$) distributions at $30\,\GeV$, and therefore to the validity of the expected axion distributions which are used in this work. The obtained axion distributions are publicly available at Ref.~\cite{zenodo}.

\section{Sensitivity for hadronic axion decays}
\label{sec:app_hadrons}
KOTO could be in principle sensitive to decays $a\to \pi^0\pi^0\eta$ or $a\to 3\pi^0$ with a subsequent $\pi^0(\eta) \to \gamma\gamma$ decay resulting in a six-cluster event. While studying hadronic decays in the kaon mode would require a dedicated analysis to address the various backgrounds, in the case of beam dump mode, it is stated in Ref.~\cite{KOTO:beamdump} that no 6-cluster events have been found in the collected sample, indicating that the $K_L$ background is kept under control in this case.

After running a simulation with {\sc Alpinist} for $a\to 3\pi^0$ and $a \to \pi^0\pi^0\eta$ decays with simple selection criteria on minimal cluster energy and cluster separation as for the $a\to \gamma\gamma$ decay in the beam dump mode, we did not find KOTO sensitivity with hadronic decays to surpass the sensitivity using $a \to \gamma\gamma$ even for larger axion masses. As can be seen in Fig.~\ref{fig:sens} for KOTO Step 2, the difference in the number of observable signal events $N_\text{sig}$ is by several orders of magnitude smaller for a combined search for $a\to 3\pi^0$ and $a \to \pi^0\pi^0\eta$ decays compared to $a \to \gamma\gamma$. Nevertheless, for convenience, we also provide the resulting datasets for these hadronic decays.

\bibliographystyle{hunsrt.bst}
\bibliography{ALPatKOTO}

\end{document}